\documentstyle[psfig]{elsart}

\newcommand{\bd}[1]{ \mbox{\boldmath $#1$}  }
\begin{document}
\begin{frontmatter}

\title{Momentum distributions and reaction mechanisms for breakup of
two--neutron halos}

\author{E. Garrido}
\address{Instituto de Estructura de la Materia, CSIC, Serrano 123, E-28006
Madrid, Spain}
\author{D.V.~Fedorov and A.S.~Jensen}
\address{Institute of Physics and Astronomy,
Aarhus University, DK-8000 Aarhus C, Denmark}
\date{\today}

\maketitle

\begin{abstract}
A theoretical model able to describe fragmentation reactions of
three--body halo nuclei on different targets, from light to heavy, is
used to compute neutron and core momentum distributions. Both Coulomb
and nuclear interactions are simultaneously included.  We specify the
different reaction mechanisms related to various processes.  The
method is applied to fragmentation of $^6$He and $^{11}$Li on C and
Pb. We find good agreement with the available experimental results.
\vspace{1pc}
\end{abstract}
\end{frontmatter}

\par\leavevmode\hbox {\it PACS:\ } 25.60.-t, 25.10.+s, 25.60.Gc

\paragraph{Introduction.}

Fragmentation reactions are one of the most powerful tools to
investigate halo nuclei \cite{rii94,han95,tan96,jon98}. Not only the
large interaction cross sections, but also the narrow momentum
distributions of the fragments, are clear evidence of the unusual
large spatial extension of such nuclei \cite{kob92,hum95,ale98,aum99}.
Their main properties are successfully described by few--body models,
where the halo nucleus is viewed as an inert core surrounded by a few
weakly bound nucleons \cite{zhu93,gar99}.  Models for one-neutron halo
breakup reactions and the corresponding momentum distributions have
been discussed \cite{das98,shy99}, but two-neutron halos clearly
require different methods. Several were developed in order to
understand the available experimental cross sections and momentum
distributions \cite{kob92,hum95,ale98,aum99}.  These
theoretical investigations fall in two independent groups. One
discusses nuclear breakup reactions and applies therefore only to
light targets \cite{gar99,ber98} whereas the other focuses on heavy
targets and considers only Coulomb dissociation
\cite{suz90,ber92,ban98,cob98}. At best Coulomb and nuclear breakup
are computed in independent models and the cross sections subsequently
simply added \cite{ber93,bar93}.

A consistent model describing two-neutron halo fragmentation on any
target, from very light to very heavy, with simultaneous treatment of
nuclear and Coulomb interactions was recently presented
\cite{gar00,gar00b}. Absolute values are computed of all possible
dissociation cross sections distinguished according to the particles
left in the final state. The two--neutron removal cross sections, core
breakup cross sections and interaction cross sections for different
targets are in good agreement with the measurements for energies above
100 MeV/nucleon. This is the only model for breakup of two-neutron
halos which uses the same two-body interactions in initial and final
states, is applicable for light, intermediate and heavy targets, and
provides all possible three-body observables. Even absolute cross
sections are calculated successfully \cite{gar00,gar00b} considering
the difficulties related to the intrinsic structures of core and
target. The precision in relative quantities like momentum
distributions is much higher.

The breakup reaction mechanism is inevitably different for light and
heavy targets and it is therefore surprising that the momentum
distributions are rather similar \cite{hum95}. To understand this
almost forgotten problem requires simultaneous inclusion of both
Coulomb and nuclear interactions which in itself is a problem of
general interest. Differential cross sections, i.e. momentum
distributions of the fragments, must be computed.  The purpose of this
letter is then to extract the reaction mechanism in two-neutron halo
breakup processes by providing evidence from model computations. The
calculated momentum distributions are especially well suited as test
observables as they depend sensitively on reaction assumptions.

The predictions vary substantially from the present
participant-spectator model to different models where the reaction
proceeds through intermediate two or three-body resonances or
continuum states.  The invariant neutron-neutron mass spectra after
breakup of $^6$He and $^{11}$Li exemplify the large differences
between reaction assumptions. Decay through two-body resonances
clearly produce the same spectra for both halo nuclei in contrast to
our model where the final state wave packet strongly depends on the
initial halo wave function \cite{gar00c}. We do not use three-body
continuum wave functions as in \cite{cob98}.  We apply in this letter
the one-participant model using optical potentials for one halo
particle-target interaction while the other interactions are treated
by the black disk model.

\paragraph{The model.}

The breakup reaction is described as a superposition of three
independent reactions, each corresponding to the interaction with the
target of one of the three constituents in the halo projectile. Thus,
in each of these three reactions only one of the constituents
(participant) interacts with the target, while the other two are mere
spectators. In the center of mass frame of the halo nucleus we obtain
for a spinless target that the differential cross sections for
absorption and elastic scattering of the participant $(i)$ by the
target $(0)$ are given by \cite{gar99}

\begin{equation}
 \frac{d^6\sigma _{abs}^{(i)}(\bd{P}^{\prime },\bd{p}_{jk}^{\prime })}
{ d\bd{P}^{\prime } d\bd{p}_{jk}^{\prime } }
=   \sigma _{abs}^{(0i)}(p_{0i}) 
 | M_s(\bd{p}_{i,jk}, \bd{p}^\prime_{jk} |^{2} 
\label{eq1}
\end{equation}

\begin{equation}
 \frac{d^9\sigma _{el}^{(i)}(\bd{P}^{\prime },
\bd{p}_{jk}^{\prime },\bd{q})}
{ d\bd{P}^{\prime } d\bd{p}_{jk}^{\prime } d\bd{q} }
  =  P_{\mbox{\scriptsize dis}}(\bd{q}) 
\frac{d^3\sigma _{el}^{(0i)}(\bd{p}_{0i}
  \rightarrow  \bd{p}_{0i}^{\prime})}{d\bd{q}} 
 | M_s(\bd{p}_{i,jk}, \bd{p}^\prime_{jk} |^{2}\nonumber
\label{eq2}
\end{equation}
where $M_s$ is the normalized overlap matrix element between the
three--body projectile and the final state spectator wave functions,
$\bd{P}^{\prime }$ is the relative momentum in the final state between
center of mass of target-participant and the spectators $j$ and $k$,
while $\bd{p}_{jk}^{\prime }$, $\bd{p}_{0i}$ and $\bd{p}_{i,jk}$ are
relative momenta between particles $j$ and $k$, $0$ and $i$, $i$ and
center of mass of $j$ and $k$. The momentum transfer is denoted by
$\bd{q}$, and primes denote final states.

Here $\sigma _{abs}^{(0i)}$ and $d^3\sigma_{el}^{(0i)}/d\bd{p}_{0i}$
are absorption and differential elastic scattering cross sections for
two-body reactions between participant and target.  These cross
sections contain nuclear and Coulomb contributions as well as the
interference between them. To remove elastic scattering reactions of
the three--body system as a whole we introduce the probability for
dissociation of the three--body system $P_{\mbox{\scriptsize
dis}}(\bd{q}) = 1-|\langle \Psi | e^{i \bd{q}_{cm} \cdot
\bd{r}_{i,jk}} |\Psi\rangle|^2$, where $\Psi$ denotes the three--body
projectile wave function and $\bd{q}_{cm} \propto \bd{q}$ is the
momentum transfer to the center of mass of the projectile.

The observable momentum distributions for any fragment are obtained by
integration of Eqs.(\ref{eq1}) and (\ref{eq2}) over the remaining
variables \cite{gar99}. When the participant is charged this
integration over $q$ diverges logarithmically for small $q$.  The
divergence disappears by removal of the virtual excitations arising at
impact parameters larger than the adiabatic distance. Since the energy
transfer also must be larger than the three--body separation energy a
minimum momentum transfer is established, i.e. $q>q_{\mbox{\scriptsize
min}}$, see \cite{gar00}.  The model then gives two qualitatively
different contributions to the measured particle momentum
distributions, i.e. one where the neutron (core) is scattered by the
target and one where the neutron-core (neutron-neutron) system
continues its motion after the instantaneous removal of the other
neutron.

\paragraph{Reaction geometries.}

\begin{figure}
\centerline{\psfig{figure=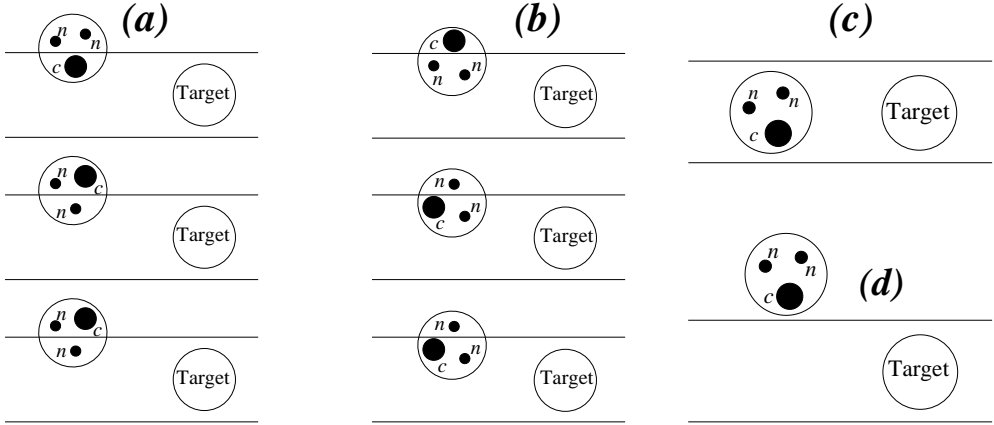,width=7cm,%
bbllx=7cm,bblly=-10.5cm,bburx=23.4cm,bbury=13.5cm,angle=0}}
\vspace*{-4cm}
\caption[]{Sketch of the possible geometries for the collision between
target and the two-neutron plus core bound three--body projectile.}
\label{fig1}
\end{figure}

The finite extension of the projectile constituents and the target is
partially destroying the simple picture described above. Simultaneous
collisions with the target of more than one constituent have to be
considered. In Fig.~\ref{fig1} we sketch the {\it geometries} needed
to describe the reaction. The short-range target-halo interactions
only act for the constituents inside the cylinder along the beam axis
around the target (Figs.~\ref{fig1}a, b and c).

When the core is inside the cylinder the core--target interaction
includes the nuclear interaction, the low impact parameter part (large
momentum transfer) of the Coulomb interaction, as well as the
interference between them. The reaction in Fig.~\ref{fig1}d  
contains the large impact parameter part (or small momentum transfer)
of the core--target interaction. It includes Coulomb elastic scattering
while the two neutrons survive untouched in the final state.
The value of the momentum transfer dividing into
low and large impact parameters is given by \cite{ber88} $q_g =
Z_0Z_ce^2 (\gamma+1)/(c \gamma \beta (R_0+R_c+ \pi a /2))$.  Here
$R_0$ and $R_c$ are charge root mean square radii of the target and
the core and $a$ is half the distance of closest core-target approach,
$eZ_0$ and $eZ_c$ are the charges of the target and core, $\beta =
v/c$ and $\gamma =1/\sqrt{1-\beta^2}$.

Let the probabilities be $P_c$ or $P_n$ respectively for finding the
core $c$ or the neutron $n$ inside the cylinders in Fig.~\ref{fig1}.
Then $P_c=P_c^{ab}+P_c^{el}$ and $P_n=P_n^{ab}+P_n^{el}$, where the
superscripts $ab$ and $el$ indicate absorption and elastic scattering
by the target. When the halo projectile hits the target each of the
reactions in Fig.~\ref{fig1}a occurs with probability
$P_i(1-P_j)(1-P_k)$, where the index $i=c,n$ refers to the constituent
inside the cylinder. Each reaction in Fig.~\ref{fig1}b occurs with the
probability $P_iP_j(1-P_k)$, where $i$ and $j$ again refer to
constituents inside the cylinder. In analogy the reaction in
Fig.~\ref{fig1}c occurs with the probability $P_iP_jP_k$. The total
probability for a process producing a specific set of halo particles
in the final state is now obtained by adding the probabilities of all
contributing processes.  Then the probabilities for two--neutron
removal (core survival) and core breakup processes are:
\begin{eqnarray}
&& P(\sigma_{-2n}) = P_c^{el} + P_n(1-P_c) + P_n(1-P_n)(1-P_c) \; ,
\label{eq3} \\ && P(\sigma_{-c} ) = P_c^{ab} \; .       \label{eq4}
\end{eqnarray}
Two--neutron removal cross sections are correspondingly obtained as a
sum of the three contributions in Eq.(\ref{eq3}).  The first term is
found by integration of Eq.(\ref{eq2}) where the participant $(i)$ is
the core and $d^3\sigma_{el}^{(0i)}/d\bd{p}_{0i}$ the differential
elastic core--target cross section containing nuclear and Coulomb
interaction and the interference between them.  Since $P_c^{el}$ is
the probability for the core being inside the cylinder only large
momentum transfer, $q>q_g$, should be included in Eq.(\ref{eq2}).
However, the two--neutron removal cross section also receives a
contribution from the reaction in Fig.~\ref{fig1}d, which contains the
low momentum transfer part ($q_{\mbox{\scriptsize min}}<q<q_g$) of the
core--target interaction.  Thus, $q$ is actually only restricted by
$q>q_{\mbox{\scriptsize min}}$.

The second term in Eq.(\ref{eq3}) corresponds to a process in which
one of the halo neutrons interact with the target while the core is
outside the cylinder. The contributions to the two--neutron removal
cross section correspond to neutron absorption and elastic scattering
obtained respectively from Eqs.(\ref{eq1}) and (\ref{eq2}), where the
core impact parameters should be larger than the radius of the
cylinder. The third term in Eq.(\ref{eq3}) corresponds to a process in
which one of the halo neutrons interact with the target while the
other neutron and the core both are outside the cylinder. The
contributions to the two--neutron removal cross section again
correspond to neutron absorption and elastic scattering obtained
respectively from Eqs.(\ref{eq1}) and (\ref{eq2}), where both impact
parameters for the core and the other neutron should be larger than
the radius of the cylinder. The core breakup cross section
corresponding to Eq.(\ref{eq4}) is simply obtained from Eq.(\ref{eq1})
where $\sigma_{abs}^{(0i)}$ is the core--target absorption cross 
section.

The neutron and core momentum distributions after two--neutron removal
and core breakup reactions are the differential cross sections
obtained from Eqs.(\ref{eq1}) and (\ref{eq2}) by leaving out the
corresponding integrations \cite{gar99}. They contain more details and
provide therefore better tests of the reaction mechanism.

\paragraph{Numerical examples.}
We apply the method to fragmentation reactions of $^6$He and $^{11}$Li
on carbon and lead, for which experimental neutron and core momentum
distributions are available.  The wave functions of the three--body
halo projectiles are obtained by solving the Faddeev equations in
coordinate space using the neutron--neutron and neutron--core
interactions specified in \cite{gar99,cob98}.  The optical potentials
for the neutron--target, $\alpha$--target and $^9$Li--target
interactions are from \cite{coo93,nol87}, where range and diffuseness
parameters for $^9$Li--target are from \cite{zah96} and the energy
dependence of the real part of the potential has been reduced to allow
for the required large beam energy variation \cite{gar00,gar00b}.

The last ingredient needed is the radius $R^{(s)}_{cut}$ of the
cylinder determining whether the spectator $s$ is interacting with the
target. We estimate $R^{(s)}_{cut}$ by equating the experimental
spectator--target absorption cross section with the absorption cross
section in the black disk model $\pi R_{cut}^{(s) 2}$. This gives
$R^{(n)}_{cut}$=3.5 fm and 7.4 fm, $R^{(^4\mbox{\scriptsize
He})}_{cut}$=4.0 fm and 7.6 fm and $R^{(^9\mbox{\scriptsize
Li})}_{cut}$=4.8 fm and 7.9 fm, where the first and second numbers
refer to C and Pb targets. These values are very close to those of
\cite{gar00,gar00b} and the momentum distributions are in any case
rather insensitive to these parameters.

The integrations of Eqs.(\ref{eq1}) and (\ref{eq2}) corresponding to
the second contribution in Eq.(\ref{eq3}) should only include core
impact parameters larger than $R^{(c)}_{cut}$.  This is in practice
approximated by including only the part of the projectile wave
function where the distance between the participant neutron and the
spectator core is larger than $R^{(c)}_{cut}$. Analogously the
integrals corresponding to the third term of Eq.(\ref{eq3}) only
include the part of the three--body wave function where the distances
between the participant neutron and the spectators, neutron and core,
are larger than $R^{(n)}_{cut}$ and $R^{(c)}_{cut}$, respectively.

\begin{figure}
\centerline{\psfig{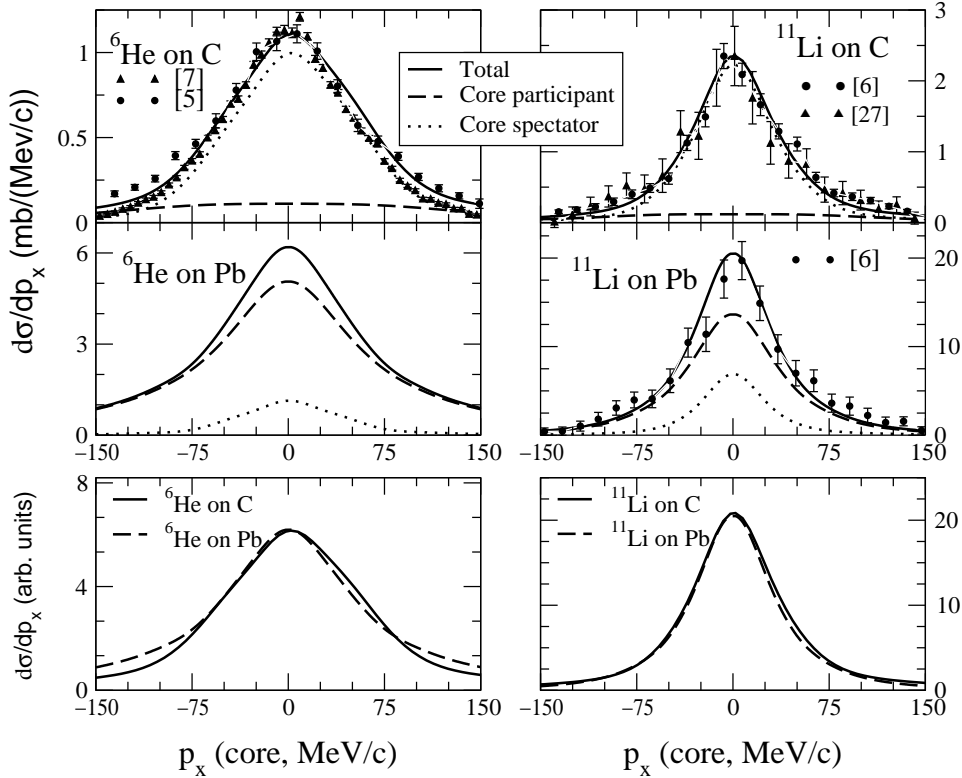}}
\vspace*{1.5cm}
\caption[]{Transverse core momentum distributions after fragmentation
of 300 MeV/nucleon beams of $^6$He and $^{11}$Li on C and Pb. Upper
part: The dashed and short--dashed lines are the contributions to the
total (solid lines) from core participant and core spectator..  The
experimental data are scaled to the calculations. They are for beam
energies between 240 and 400 MeV/nucleon where the computed
distributions are almost energy independent \cite{gar99}. The data are
from \cite{kob92,ale98} for $^6$He on C, from \cite{hum95,gei} for
$^{11}$Li on C and from \cite{hum95} for $^{11}$Li on Pb.  Lower part:
The distributions for $^6$He on C and Pb (left) and for $^{11}$Li on C
and Pb (right).}
\label{fig2}
\end{figure}

In Fig.~\ref{fig2} we show core momentum distributions after
fragmentation of $^6$He and $^{11}$Li on C and Pb.  Longitudinal and
transverse distributions are distiguishable but very similar
\cite{gar99,gar98}.  For heavy targets the main contribution comes from
the core participant and the Coulomb interaction between core and
target is all-decisive.  The dominant reaction is that of
Fig.~\ref{fig1}d.  For light targets the situation is the opposite and
the core spectator contribution dominates. The core--target
interaction plays a minor role and the dominant reactions are those of
Figs.~\ref{fig1}a and b with the core as a spectator.

It is significant that even though the momentum distributions from
light and heavy targets are related to projectile--target interactions
of different nature (nuclear neutron--target interaction and Coulomb
core--target interaction) the shape of the distributions are very
similar as seen in the lower part of Fig.~\ref{fig2}, where the
distributions are scaled to the same maximum.  The reason is that
removal of a halo neutron by a light target essentially leaves the
spectators undisturbed. The resulting momentum distribution of the
core is therefore close to the initial distribution inside the
projectile, except for the final neutron--core interaction, which only
has a relatively small effect on the core momentum \cite{gar97}. For
heavy targets the main contribution comes from the low momentum
transfer part of the Coulomb interaction.  Thus the momentum
transferred to the core in the collision is small compared to its
initial momentum distribution which therefore for an entirely
different reason again is left relatively unchanged. This explanation
is supported by the fairly good agreement between measurements and
calculations.

\begin{figure}
\centerline{\psfig{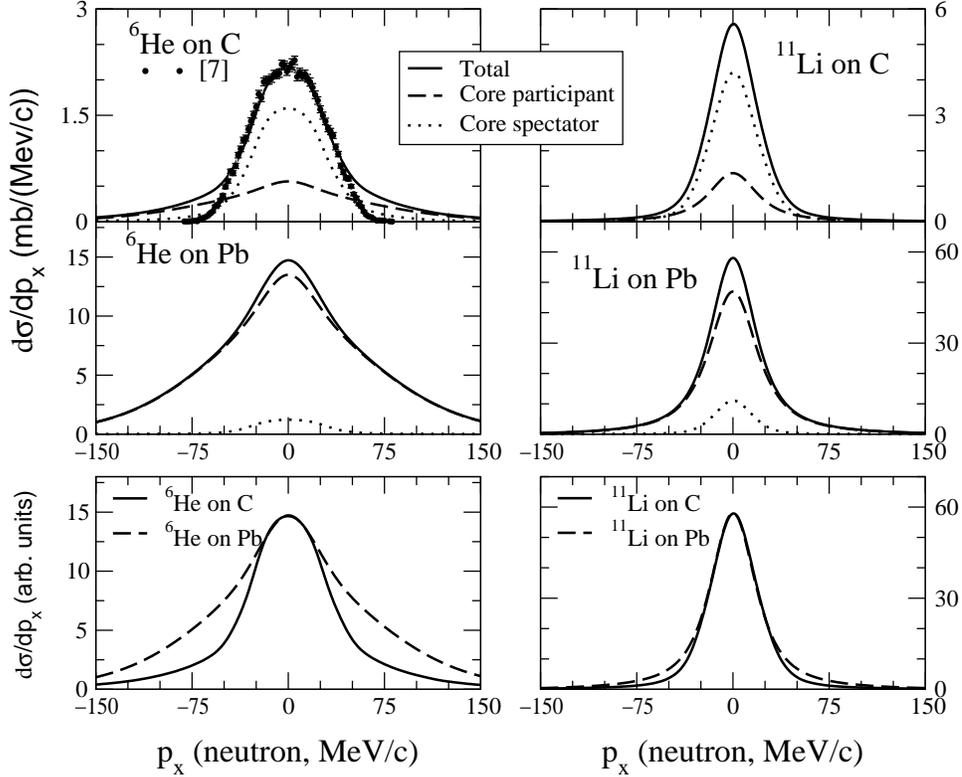}}
\vspace*{1.5cm}
\caption[]{The same as Fig.~\ref{fig2} for transverse neutron momentum
distributions.  Experimental data are from \cite{ale98} for $^6$He 
on C.}
\label{fig3}
\end{figure}

In Fig.~\ref{fig3} we show transverse neutron momentum distributions
after two--neutron removal of $^6$He and $^{11}$Li on C and Pb.  The
longitudinal distributions are hardly distinguishible from the ones
shown in the figure.  For light targets the core spectator
contribution is again dominating, while for heavy targets the main
contribution comes from processes where the core participates.  The
agreement with experiments is excellent for the carbon target. The
computed curve is above the experimental distribution in the tails due
to an experimental neutron acceptance of 50 MeV/c in both horizontal
and vertical directions.

For a lead target, although experimental neutron distributions are
not available, a discrepancy between computations and experiments
is expected.  The reason is that the neutrons,
unlike the core, are highly influenced by the final state interaction
between fragments \cite{gar97}.  In our calculations the final state
interaction is properly included between the two spectator
constituents as appropriate when one neutron is absorbed. On the other
hand when the low momentum transfer contribution dominates
(Fig.~\ref{fig1}d) with the core as participant, as for heavy targets,
the included final state interaction between the two neutrons is
insufficient to describe the data. The final state interaction between
all three halo constituents should be included and in particular the
neutron-core interaction. This implies use of correct three--body
continuum wave functions as the final state. It is in this context
worth emphasizing that such a final state would be wrong for other
reaction mechanisms responsible for processes like neutron absorption.

The neutron momentum distributions for $^{11}$Li on C and Pb targets
are very similar in the computation in contrast to the case of $^6$He
as a projectile, see the lower part of Fig.~\ref{fig3}.  This is an
interesting coincidence.  For the lead and carbon targets we include
respectively the neutron--neutron and neutron--$^9$Li final state
interaction in the dominating contributions. In both cases a low lying
virtual s--state at an energy of around 100 keV is present.  Since the
neutron--$^4$He interaction does not have such a low lying s--state
the neutron momentum distributions for $^4$He on C and Pb targets differ
substantially.  For lead the core is the participant and for carbon
the neutron is the participant. The reaction mechanisms are completely
different yet the distributions are similar when the final state
interactions are similar.  This is a powerful
illustration of the importance of the final state interaction.

\paragraph{Summary and conclusions}

Fragmentation reactions of two--neutron halo nuclei are described as
superposition of all possible reactions where one, two or three halo
constituents interact with the target. Nuclear and Coulomb
interactions are simultaneously considered allowing light and heavy
targets. The two--neutron removal and core breakup cross sections can
be described through processes where only one of the halo constituents
interact with the target. The appropriate interactions are described
by an optical potential that takes into account both elastic
scattering and absorption by the target. Initial and final state
interactions are identical.

Application of the method to fragmentation of $^6$He and $^{11}$Li
reveals that the reactions for light or heavy targets are dominated by
processes where the core is spectator or participant, respectively.
For heavy targets the core--target Coulomb interaction is
all-decisive. For targets of intermediate masses the Coulomb and
nuclear interactions and the core spectator and participant
contributions can be anticipated to compete in importance.  Core
momentum distributions on light and heavy targets are very similar in
shape even though the reaction mechanisms are completely
different. All computed neutron and core momentum distributions agree
very well with the available experiments. The neutron-core interaction
probably is essential for the neutron distribution for a heavy target, 
and the continuum three-body final state wave function is approriate.

In conclusion the reaction mechanisms described in our fully
consistent model are able to reproduce experimental total two--neutron
removal and interaction cross sections \cite{gar00,gar00b} as well as
core and neutron momentum distributions for two-neutron halos
colliding with light and heavy targets.

\paragraph*{Acknowledgement.} We thank K. Riisager for continuous 
discussions and suggestions.

\end{document}